\documentclass[showpacs,aps,pra,twocolumn,citeautoscript]{revtex4-1}

\usepackage{amsmath,amssymb}
\usepackage{graphicx}
\usepackage{times}
\usepackage{color}

\newcommand{\bfk}{{\mathbf{k}}}

\newcommand{\varA}{{\mathcal{A}}}
\newcommand{\varG}{{\mathcal{G}}}
\newcommand{\varH}{{\mathcal{H}}}

\newcommand{\ket}[1]{\left|#1\right\rangle}
\newcommand{\bra}[1]{\left\langle#1\right|}
\newcommand{\Braket}[1]{\mathinner{\langle{\textstyle#1}\rangle}}
\newcommand{\avg}[1]{\mathinner{\langle{\textstyle#1}\rangle}}

\newcommand{\eqnref}[1]{Eq.~(\ref{#1})}

\newcommand{\figref}[1]{Fig.~\ref{#1}}
\newcommand{\figsref}[1]{Figs.~\ref{#1}}
\newcommand{\Figref}[1]{Figure~\ref{#1}}

\let\up\uparrow
\let\down\downarrow

\begin{document}

\title{Thermoelectric effect in the Kondo dot side-coupled to a Majorana
  fermion}

\author{Heunghwan Khim}
\affiliation{%
  Department of Physics, Korea University, Seoul 136-701, Korea%
}

\author{Rosa L\'opez}
\affiliation{%
  Institut de F\'{i}sica Interdisciplinar i de Sistemes Complexos IFISC
  (CSIC-UIB), E-07122 Palma de Mallorca, Spain%
}
\affiliation{%
  Departament de F\'{i}sica, Universitat de les Illes Balears,
  E-07122 Palma de Mallorca, Spain%
}

\author{Jong Soo Lim}
\affiliation{%
  School of Physics, Korea Institute for Advanced Study, Seoul 130-722, Korea%
}

\author{Minchul Lee}
\email{minchul.lee@khu.ac.kr}
\affiliation{%
  Department of Applied Physics, College of Applied Science,
  Kyung Hee University, Yongin 446-701, Korea%
}

\pacs{
  72.15.Jf, 
  73.63.Kv, 
  72.10.Fk  
}

\date{\today}

\begin{abstract}
  We investigate the linear thermoelectric response of an interacting quantum
  dot side-coupled by one of two Majorana fermions (MFs) formed at the ends of
  a topological superconducting wire. We employ the numerical renormalization
  group technique to obtain the thermoelectrical conductance $L$ as well as the
  electrical conductance $G$ when the background temperature $T$ and the dot
  gate are tuned. We distinguish two transport regimes in which $L$ displays
  different features: the weak- $(\Gamma_{m} < T_{K})$
  and strong-coupling $(\Gamma_{m} > T_{K})$ regimes, where $\Gamma_{m}$ and
  $T_{K}$ are the Majorana-dot coupling and the Kondo temperature,
  respectively.
  For an ideal (infinitely long) nanowire where the Majorana end states do not
  overlap $(\epsilon_{m} = 0)$, the thermoelectrical conductance $L$ in the
  weak-coupling regime exhibits a peak at $T \sim \Gamma_{m}$. This peak is
  ascribed to the anti-Fano resonance between the asymmetric Kondo resonance
  and the zero-energy MF mode. Interestingly, in the strong-coupling regime,
  the Kondo-induced peak in $L$ is shifted due to the MF-induced Zeeman
  splitting in the dot.
  For finite but small $\epsilon_{m} > 0$, the interference between two MFs
  restores the Kondo effect in the dot in a smaller energy scale
  $\Gamma^{\prime}_{m}$ and gives rise to an additional peak in $L$ at $T \sim
  \Gamma^{\prime}_{m}$, whose sign is opposite to that at $T\sim\Gamma_{m}$. In
  the strong-coupling regime this additional peak can cause a non-monotonic
  behavior of $L$ with respect to the dot gate.
  Finally, we examine the case in which an ordinary spin-polarized fermion is
  coupled to the dot and identify the fingerprint of MFs by comparing two
  cases.
\end{abstract}

\maketitle

\section{Introduction}
\label{sec:introduction}

The advent of topological materials \cite{Hasan2010} has ignited a tremendous
interest of the realization of quantum operations in platforms immune to
decoherence processes \cite{Nayak08}. A prominent feature of such materials is
that they exhibit quasiparticle excitations similar to the elementary particles
predicted by Ettore Majorana \cite{Ettore37} and named Majorana fermions (MFs)
\cite{Kitaev01,Wilczek09,Alicea12,Beenakker13}.  In particular, the exoticness
of such particles is that they coincide with their own anti-particles. The
enormous interest generated by Majorana quasiparticles in solid state systems
resides in the possibility of creating such excitations non-locally and
manipulate them for quantum computation purposes \cite{Kitaev01}. Majorana
quasiparticles behave as nonlocal qubits being resistant to decoherence
phenomena. By exchanging two Majorana quasiparticles a non-trivial quantum
operation (unitary transformation or braiding operation), is performed. Indeed,
braiding manipulations among Majorana quasiparticles are the fundamental basis
for the realization of topological quantum computation \cite{Nayak08}.

There are a plethora of proposals for the observation of Majorana
quasiparticles in a diverse of solid state setups: quantum Hall states,
$p$-superconductors, or topological insulators
\cite{Read91,Kitaev01,Ivanov01,Sarma06,Ku08,Linder10a} among others.
Nevertheless, the first signatures of the occurrence of such exotic excitations
was reported in quasi-one-dimensional semiconductor nanowires
\cite{Mourik12}. This system consists of a nanowire with strong spin-orbit
interaction put in proximity to a superconductor, and exposed to a magnetic
field
\cite{Yuval10,Alicea10,Lutchyn10,Linder10b,Potter11,Liu12,Pientka12,Elsa12a,Lim12}.
The experimental evidence of such quasiparticles
\cite{Mourik12,Deng12,Heiblum12, Churchill12,Finck13} was realized by means of
tunnel spectroscopy. However, so far, these evidences are not totally
conclusive.  In these experiments, the appearance of a zero-bias anomaly (ZBA)
in the nonlinear conductance is ascribed to the Majorana physics. Nevertheless,
recently it has been pointed out that other sources of ZBA in
normal-superconductor nanowires can occur.  For instance, Kondo physics or
Shiba states \cite{Chang13,Eduardo12,Rok14} or even nearly zero energy Andreev
states \cite{Kells12,Eduardo13} and weak antilocalization \cite{Pikulin12a}
cannot be discarded as explanations of the observed ZBAs. Therefore, the
detection of the MFs is now a dire challenge. Most of schemes for the detection
up to now are exploiting the electronic transport involving the MFs. And only
very recently the thermoelectric properties of Majorana fermion are examined
and suggested for another way to distinguish Majorana physics from other
explanations \cite{Lopez:2014,Leijnse:2014}. Since the MF has the particle-hole
symmetry itself, thermoelectric devices involving only the MF cannot generate
any temperature-driven charge current nor electric-field-driven heat
current. Hence, an additional quantum object which breaks the particle-hole
symmetry should be incorporated. The recent studies
\cite{Lopez:2014,Leijnse:2014} have coupled the MF to the simplest quantum
object, that is, a spinless quantum dot (QD) and controlled the particle-hole
symmetry by tuning the dot level. It turns out that the MF can give rise to a
non-trivial thermoelectric response upon a thermal bias across the QD
\cite{Lopez:2014}.

Interests for the reciprocity between heat and charge currents has been revived
in systems at nanoscale \cite{Dhar08,Dubi11}. The electric response to a
temperature gradient, in nanostructures, can achieve much bigger values in
efficiency compared with macroscopic samples
\cite{Butcher90,Molenkamp92,Dzurak97,Godjin99}. The main reason for such high
heat-to-electricity conversion factors resides in the intrinsic quantum
property of energy level discretization for confined systems
\cite{Molenkamp92}. However, in a more fundamental basis, thermoelectric
transport can reveal rather useful information about the intrinsic nature of a
quantum system \cite{Coleman05,Jacquod10,Balachandran12} and about their
interactions \cite{pekola08}.  In this respect, our investigation explores the
thermoelectric response to characterize the half-fermionic nature of Majorana
nanowires.

\begin{figure}[!b]
  \includegraphics[width=9cm]{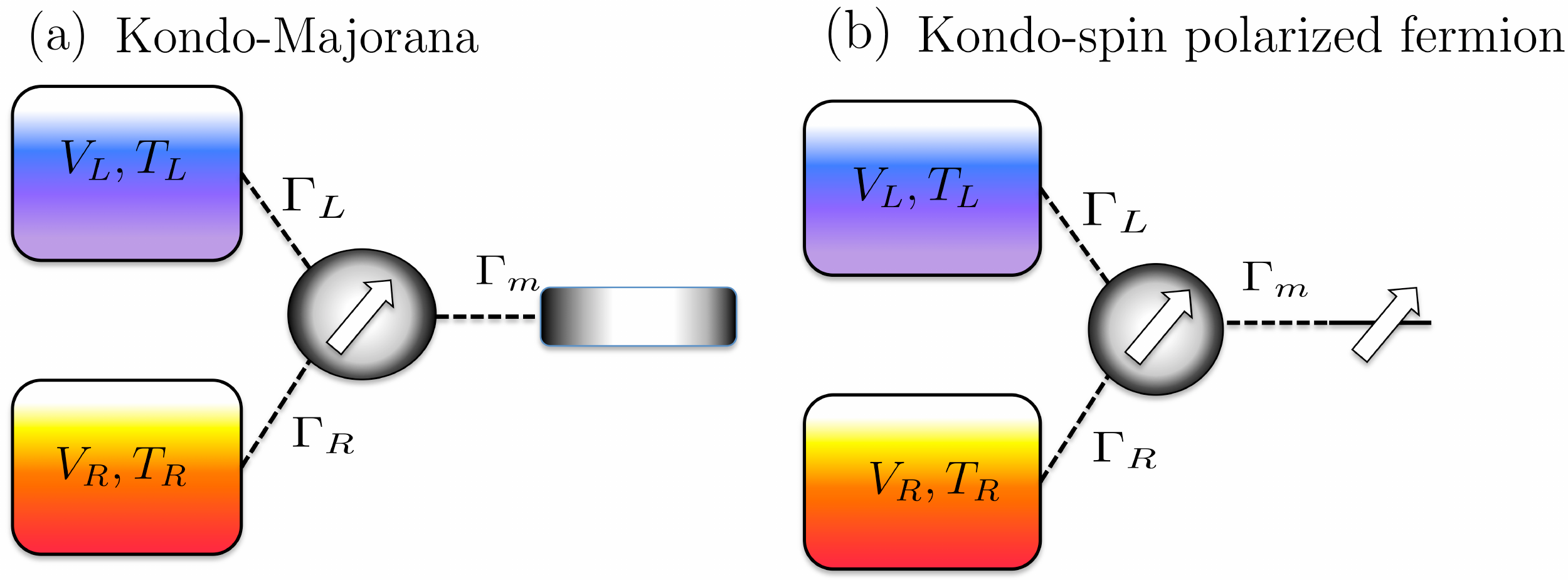}
  \caption{(Color online) Quantum dot system coupled to two normal-metal leads
    (by tunneling rates $\Gamma_{L(R)}$) and to (a) the MF state formed at one
    end of a topological superconductor wire and (b) the spin-polarized (SF)
    ordinary fermion state by QD-MF or QD-SF tunneling rate $\Gamma_m$. Due to
    the helical property of the MF state, only one of the spin components (say
    spin-$\down$) is coupled to the MF. In the SF case, the spin is assumed to
    be polarized in the spin-$\down$ direction.}
  \label{fig:model}
\end{figure}

In the context of the MF, mostly the spinless or spin-polarized QD is chosen
because an external magnetic field is necessary for inducing the MF. However,
as long as the Zeeman splitting in the QD is small enough or the external
magnetic field is replaced by a proximity induced effective field due to a
ferromagnetic layer, one can consider the case of the spinful QD in which the
Coulomb interaction is highly relevant. In fact, the interplay of the Coulomb
interaction and the MF in the Kondo dot side-coupled to a MF [see
\figref{fig:model}(a)] is found to display an interesting transport behavior
\cite{Minchul13}: The presence of the MF modifies the Kondo effect profoundly
by reducing the linear conductance by a factor 3/4 or by inducing an effective
Zeeman splitting on the QD, which leads to spin-split Kondo resonance.


In Ref. [\onlinecite{Minchul13}] some of the authors of this work investigated
the interplay of Majorana and Kondo physics in the electrical
conductance. Here, the purpose is to study the thermoelectric response of the
side-coupled QD-nanowire system [see \figref{fig:model}(a)] as a tool to detect
Majorana quasiparticles. A temperature difference is established between the
two normal contacts coupled to the QD, then the electric response is measured.
We find that thermoelectric properties in the Kondo dot-MF system depend
strongly on the QD-MF coupling ($\Gamma_{m}$), the overlap between two MF
end-states ($\epsilon_{m}$), and the Kondo temperature $(T_{K})$. For the ideal
case ($\epsilon_{m} = 0$), due to the anti-Fano resonance between Kondo
resonance peak and the zero-energy MF mode, the thermoelectrical conductance
has a finite value at $T \sim \Gamma_{m}$ in the weak-coupling regime
($\Gamma_{m} < T_{K}$). In the strong-coupling regime ($\Gamma_{m}>T_{K}$), the
induced Zeeman splitting shifts the Kondo-induced peak in the thermoelectrical
conductance. The finite overlap ($0<\epsilon_{m}<\Gamma_{m}$) leads to the
interference between Majorana fermions and restores the Kondo effect in a small
energy scale $\Gamma'_m$ and the thermoelectrical conductance becomes finite at
$T\sim\Gamma'_{m}$ whose sign is opposite to that at $T\sim\Gamma_m$. We also
consider the case in which the MF is replaced by an ordinary spin-polarized
fermion (SF) and find similarities and differences between two cases,
eventually identifying the fingerprint of MF in the thermoelectric effect.

In Sec. II we introduce the model of Hamiltonian employed to describe the QD-MF
setup and the QD side-coupled to a spin polarized fermion.  Electrical and
thermoelectrical conductances are also derived in terms of the local spectral
density for the interacting QD.
Sec. III and IV explain our results for the temperature dependence of the
electrical and thermoelectrical conductances for the QD-MF device and a
comparison with the spin polarized fermion model is shown. Section V is devoted
to describe our findings for the gate dependence of the thermoelectrical
conductance, and we finish by enumerating the main conclusions in Sec. VI.

\section{Model and Methods}

\subsection{Model Hamiltonian for quantum dot system side-coupled to Majorana fermion}

We consider a two-terminal system in which a QD is tunnel-coupled to two
normal-metal leads and to one end of a topological superconducting wire (TSW)
which hosts two MFs at its ends [see \figref{fig:model}(a)]. The on-site
Coulomb repulsion in the QD is considered. The voltage bias and/or the thermal
gradient will be applied across the QD. Assuming that the superconducting gap
in the TSW is large enough, the effective low-energy Hamiltonian reads
\begin{align}
  \label{eq:H:MF}
  \begin{split}
    \varH
    & =
    \sum_{\ell\bfk\mu} \epsilon_\bfk c^{\dagger}_{\ell\bfk\mu} c_{\ell\bfk\mu}
    + \sum_{\ell\bfk\mu}
    \left[t_{\ell} d^{\dagger}_{\mu} c_{\ell\bfk\mu} + (h.c)\right]
    \\
    & \quad\mbox{}
    + \epsilon_{d} \sum_{\mu} n_\mu + U n_\up n_\down
    \\
    & \quad\mbox{}
    + 2i \epsilon_{m}\gamma_{1} \gamma_{2}
    + t_{m} (d^{\dagger}_\down\gamma_{1} + \gamma_{1}d_\down).
  \end{split}
\end{align}
Here $c^{\dagger}_{\ell\bfk\mu}$ creates an electron with momentum $\bfk$,
energy $\epsilon_{\bfk}$, and spin $\mu$ in lead $\ell=L, R$. Two leads are
assumed to share a same flat-band dispersion $\epsilon_\bfk$ with a half
bandwidth $D$ and density of states $\rho$.
The operator $d^{\dagger}_{\mu}$ creates an electron in the single-level QD
with energy $\epsilon_{d}$ and spin $\mu$: $n_{\mu} = d^{\dagger}_{\mu}
d_{\mu}$ is the dot occupation for spin $\mu$. The Coulomb repulsion in the QD
is denoted by $U$. As mentioned in Sec.~\ref{sec:introduction} the QD is
considered to be spin degenerate \cite{Leijnse:2011, Minchul13}. This condition
can be achieved by using low $g$ factor material for the QD or by using
proximity with a magnetic insulator to induce the effective Zeeman splitting in
the TSW \cite{Fu:2008,Sau:2010}.
The tunneling between the leads and the QD is quantified by the tunneling
amplitude $t_\ell$, which defines the tunneling rates $\Gamma_\ell = \pi \rho
|t_\ell|^2$. For simplicity, the tunneling rates are assumed to be
momentum-independent, and the total lead-QD tunneling rate is denoted as
$\Gamma = \Gamma_L + \Gamma_R$.

The low-energy Hamiltonian of the TSW is governed by the two MF operators
$\gamma_i = \gamma_i^\dag$ ($i=1,2$) \cite{Kitaev:2001} which satisfy the
Clifford algebra $\{\gamma_{i},\gamma_{j}\} = \delta_{ij}$. In a finite-length
TSW the two MF states have a finite overlap between their wave functions, which
is quantified by the overlap integral $\epsilon_m$. In addition, the MF
operators can be written in terms of an ordinary fermion operator $f$ which
satisfies the usual fermionic commutation relation $\{f,f^{\dagger}\} = 1$ as
\begin{align}
  \gamma_{1} = \frac{f+f^{\dagger}}{\sqrt{2}},
  \qquad
  \gamma_{2} = \frac{f-f^{\dagger}}{i\sqrt{2}}.
\end{align}
In terms of this $f$ operator, the Majorana coupling term in \eqnref{eq:H:MF}
can be rewritten as $2i\epsilon_m \gamma_1 \gamma_2 = \epsilon_m (f^\dag f -
1/2)$, which means that $\epsilon_m$ is nothing but the energy splitting
between empty ($\ket0$) and filled ($\ket1 = f^\dagger\ket0$) states.
We assume that the QD is coupled to the nearest MF (say $\gamma_1$) and that
the coupling to the other MF is negligible. The QD-MF tunneling amplitude $t_m$
characterizes the tunneling rate $\Gamma_{m} = \pi \rho_{\textrm{dot}}
|t_{m}|^2$, where $\rho_{\rm dot}$ is the density of states in the QD at the
Fermi level.  It should be noted that due to the helical property of the MF
states, only one of the QD spin orientations (say spin-$\down$) hybridizes with
the MFs \cite{Flensberg:2010,Sticlet:2012,Minchul13}.

\subsection{Model Hamiltonian for quantum dot system side-coupled to ordinary spin-polarized fermion}

In order to clarify the genuine effect by the MF it is useful to compare the MF
system with the non-MF system having the same configuration. Explicitly, we
replace the MF operator in \eqnref{eq:H:MF} by an ordinary spin-polarized
fermion operator $f_\down$ [see \figref{fig:model}(b)].
%
In this SF case, the last line of the Hamiltonian \eqnref{eq:H:MF} is changed
into
\begin{align}
  \label{eq:H:SF}
  \epsilon_m f_\down^\dag f_\down
  + t_m (d_\down^\dag f_\down + f_\down^\dag d_\down).
\end{align}
In fact, the first term is effectively same as the MF coupling term (except a
constant term) if $f_\down$ operator is replaced by $f$ operator. The
difference between two cases arises in the tunneling term. In the SF case, the
second term in \eqnref{eq:H:SF} represents the hopping of an electron between
the QD and the SF level. In the MF case, the tunneling term
\begin{align}
  \frac{t_m}{\sqrt2} [d_\down^\dag (f+f^\dag) + (f+f^\dag) d_\down]
\end{align}
includes not only the electronic hopping but also the Cooper pair
creation/annihilation. This superconducting coupling should distinguish the MF
case from the SF case.

\subsection{Linear response: electrical and thermoelectrical conductances}

We are interested in measuring the charge current through the QD upon the
voltage bias and/or temperature gradient between two terminals [see
\figref{fig:model}]. The current is the expectation value of the charge current
operator which is given by the time derivative of the dot occupation
$\sum_{\mu} n_\mu$. Using the equation-of-motion technique, the charge current
can be expressed in terms of the retarded QD Green's function
$\varG^R_\mu(t,t') = -i\Theta(t-t^{\prime})
\Braket{\{d_{\mu}({t}),d^{\dagger}_{\mu}(t^{\prime})\}}$ as
\begin{align}
  I
  =
  \frac{4e}{h}
  \frac{\Gamma_{L}\Gamma_{R}}{\Gamma}
  \int{d\epsilon}
  \left(f_L(\epsilon) - f_R(\epsilon)\right)
  \sum_{\mu} \pi\varA_\mu(\epsilon),
\end{align}
where $f_\ell(\epsilon) = (1 + e^{(\epsilon-eV_\ell)/k_BT_\ell})^{-1}$ is the
Fermi-Dirac distribution with the voltage $V_\ell$ and the temperature $T_\ell$
and $\varA_\mu(\epsilon) = - {\rm Im} \varG^R_\mu(\epsilon)/\pi$ is the QD
spectral density.
Assuming that the temperature gradient $\delta T \equiv T_{L} - T_{R}$ and the
voltage bias $\delta V \equiv V_{L}-V_{R}$ across the leads are small enough,
the charge current can be linearized as
\begin{align}
  I= G \delta V + L \delta T,
\end{align}
where the electrical conductance $G$ and the thermoelectrical conductance $L$
are given by
\begin{subequations}
  \begin{align}
    G
    & =
    \frac{e^{2}}{h}
    \frac{4\Gamma_{L}\Gamma_{R}}{\Gamma}
    \int{d\epsilon}
    \left(-\frac{\partial f_{\rm eq}(\epsilon)}{\partial \epsilon}\right)
    \sum_\mu \pi\varA_\mu(\epsilon),
    \\
    \nonumber
    L
    & =
    \frac{ek_B}{h}
    \frac{4\Gamma_{L}\Gamma_{R}}{\Gamma}
    \int{d\epsilon}
    \frac{\epsilon - \epsilon_F}{k_BT}
    \\
    & \qquad\qquad\qquad\mbox{}
    \times
    \left(-\frac{\partial f_{\rm eq}(\epsilon)}{\partial \epsilon}\right)
    \sum_\mu \pi\varA_\mu(\epsilon).
  \end{align}
\end{subequations}
Here $f_{\rm eq}(\epsilon)$ is the equilibrium Fermi-Dirac distribution
function at the Fermi energy $\epsilon_F$ and the background temperature
$T$. In the low temperature limit the two linear conductances can be
analytically obtained by using a Sommerfeld expansion, then
\begin{subequations}
  \begin{align}
    G
    & \approx
    \frac{e^2}{h} \mathcal{T}_d(\omega)|_{\omega=\epsilon_F},
    \\
    L
    & \approx
    -\frac{k_B e}{h} \frac{\pi^2 k_B T}{3}
    \left.
      \frac{\partial_{\omega}T_{d}(\omega)}{T_{d}(\omega)}
    \right|_{\omega=\epsilon_F},
  \end{align}
\end{subequations}
where $\mathcal T_d(\omega) = (4\Gamma_L\Gamma_R/\Gamma) \sum_\mu \pi
A_{\mu}(\omega)$ is the dot transmission.  It should be noted that $L$ becomes
finite only if the spectral density is asymmetric with respect to the Fermi
level: if the spectral density has larger weight at $|\omega| \sim T$ in the
particle (hole) part than in the hole (particle) part, $L$ (in unit of
$k_Be/h$) becomes positive (negative) at temperature $T$.

\subsection{NRG and Kondo regime}

The NRG method \cite{wilson:1975, Krishna:1980} is known to be highly effective
in studying the effect of the strong Coulomb interaction nonperturbatively. We
have already used the NRG technique for investigating the electric property of
our system in Ref.~[\onlinecite{Minchul13}], finding very interesting interplay
between the Majorana and Kondo physics. For a full review of the NRG, one can
refer to Ref.~[\onlinecite{Bulla:2008}]. Below we give a brief introduction of
the NRG used in this work.

The NRG method consists in a series of consecutive exact diagonalizations of an
appropriately discretized form of $\varH$ represented by a linear tight-binding
chain.  In each partial diagonalization step the spectrum is truncated by
retaining only those low-energy excitations that contains the strongest coupling
terms.  Better improvements of the NRG approach such as the density-matrix NRG
\cite{Hofstetter:2000}, the $z$-averaging, and refinement of the logarithmic
discretization \cite{campo:2005, zitko:2009} allows us to calculate the
dynamical properties at higher energies in a reliable and rather accurate way,
which is essential for our thermoelectric study.
The NRG procedure also exploits the existence of symmetries for better
efficiency. Note that this system has two conserved quantities: $[Q_\up,\varH]
= [P_\down,\varH] = 0$ where $Q_\up$ and $P_\down$ are charge number operator
for spin-$\up$ electrons and parity operator for the sum $Q_\down$ of
spin-$\down$ electrons and $f$ electrons, respectively. Note that the QD-MFS
hopping changes $Q_\down$ by even numbers only.
To obtain the electrical conductance $G$ and the thermoelectrical conductance
$L$, at each iteration in the backward stage, the spin-resolved QD spectral
density $\varA_{\mu}(\epsilon)$ is calculated
\begin{align}
  \begin{split}
    \varA_{\mu}^{(N)}(\epsilon)
    & =
    \sum_{nm} \Braket{n|d^{\dagger}_{\sigma}|m}_N
    \Braket{m|d_{\sigma} \rho^{\rm red}_N + \rho^{\rm red}_N d_{\sigma}|n}_N
    \\
    & \qquad\qquad\mbox{}
    \times \delta[\epsilon - (E_{n}^{(N)}-E_{m}^{(N)})],
  \end{split}
\end{align}
where $\ket{n}_N$ and $E_{n}^{(N)}$ are the eigenstates and the corresponding
eigenenergies obtained at iteration $N$. The reduced density matrix $\rho^{\rm
  red}_N$ is iteratively calculated in backward process, starting at the final
iteration $M$ which is determined by the condition that $T \sim T_M$ where
$T_M$ is the energy scale corresponding to iteration $M$. At the final
iteration, the reduced matrix is given by
\cite{Bulla:2001,Hofstetter:2000,Bulla:2008}
\begin{align}
  \rho^{\rm red}_M
  =
  \sum_n \frac{e^{-\beta E_{n}^{(M)}}}{Z_M} \ket{n}_M \bra{n}
  \quad\text{with}\quad
  Z_M = \sum_n e^{-\beta E_{n}^{(M)}}.
\end{align}
Since we are interested in the Kondo regime, throughout our work, we have used
the following values of the parameters unless specified otherwise:
$\epsilon_{d}=-0.2D$, $U=D$, and $\Gamma=0.04D$ using the bandwidth $D$ as the
energy unit ($D=1$).

\section{Long TSW Case: $\epsilon_{m}=0$}

For an ideal TSW the two MF end states have zero overlap. The parameter that
quantifies the degree of overlap between the two end states is denoted by
$\epsilon_m$ that is considered null through this section.  Here, we present
our NRG results for both electrical and thermoelectrical conductances in this
case. Indeed, the electric response in the QD-MF setup was discussed in
Ref.~[\onlinecite{Minchul13}], however for completeness we briefly explain the
main findings for the electrical conductance and discuss in detail the
thermoelectrical response of our device. Importantly, as we discuss below, $L$
exhibits different behaviors for the weak- ($\Gamma_m<T_K$) and strong-
($\Gamma_m>T_K$) coupling regimes.

\subsection{Weak-coupling regime, $\Gamma_{m} < T_{K}$}

\begin{figure}
  \includegraphics[width=8.5cm]{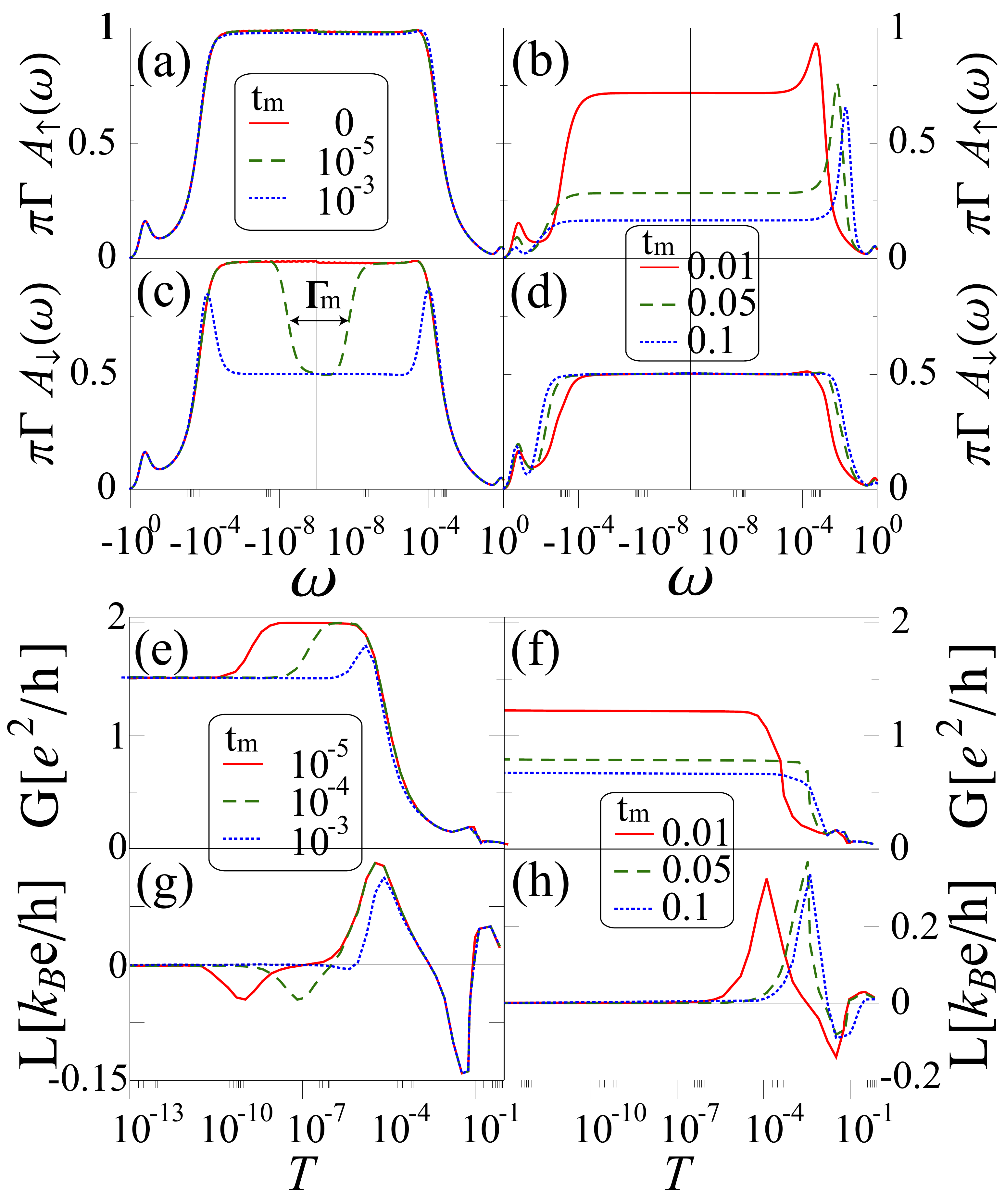}
  \caption{(Color online) Dynamical and electric/thermoelectric transport
    features in the MF case for $\epsilon_m = 0$. Left and right panels
    correspond to the weak- ($\Gamma_m<T_K$) and strong-coupling
    ($\Gamma_m>T_K$) regimes, respectively: [(a),(b)] QD spin-$\up$ spectral
    density, [(c),(d)] QD spin-$\down$ spectral density, [(e),(f)] electrical
    conductance, and [(g),(h)] thermoelectrical conductance for annotated
    values of $t_m$. Here we have used $\delta=0.8$ ($\epsilon_{d}=-0.2$ and
    $U=1$) and $\Gamma=0.04$. $T_{K}=2.64\times10^{-4}$ with these parameters.}
  \label{fig:MF:em0}
\end{figure}

First, we investigate the weak-coupling regime ($\Gamma_m < T_K$) [see the left
panels of \figref{fig:MF:em0}]. When the QD is uncoupled from the TSW, i.e.,
when $t_m = 0$, the QD spin-resolved spectral densities $A_\mu(\omega)$ [see
\figref{fig:MF:em0}(a) and \ref{fig:MF:em0}(c)] exhibit a Kondo resonance peak
located at $\omega = 0$ with a half-width $T_K$ and two side resonance peaks at
$\omega \approx \epsilon_d,\epsilon_d+U$. For a finite but weak QD-MF coupling
($t_m>0$), while $A_\up(\omega)$ is not affected by the coupling to the MF
since the spin $\up$ is not directly coupled to the MF [see
\figref{fig:MF:em0}(a) and \ref{fig:MF:em0}(c), for $t_m= 0, 10^{-3}$,
$10^{-5}$], $A_\down(\omega)$ exhibits a half-dip whose width is comparable to
$\Gamma_{m}$ [see \figref{fig:MF:em0}(c)]. The half-fermionic dip is ascribed
to the destructive interference between the Kondo resonance and the MF states.
As a result an anti-Fano resonance occurs due to the half-fermionic nature of
the MF state \cite{Liu:2011,Minchul13}.
The resultant linear conductance $G$ as a function of the temperature is shown
in \figref{fig:MF:em0}(e).  At low temperatures ($T < \Gamma_m$), $G = e^2/h +
e^2/2h = 3e^2/2h$, owing to the half-fermionic anti-Fano resonance. For larger
temperature ($\Gamma_m < T < T_K$) the conductance restores the Kondo
unitary-limit value, $2e^{2}/h$, eventually decreasing to $0$ for $T_{K} < T$
because of the Coulomb blockade and the destruction of the Kondo effect.

The thermoelectrical conductance $L$ displays more interesting features as
shown in \figref{fig:MF:em0}(g). Before addressing the thermoelectric effect
ascribed to the Majorana physics, we first need to understand the behavior of
$L$ due to the Kondo effect itself. In the symmetric case, when $\delta=0$,
with $\delta=2\epsilon_d+U$ electron-hole symmetry holds and the
thermoelectrical conductance vanishes completely. Away from this special situation
($\delta\neq 0$) the Kondo resonance peak in $A_\mu(\omega)$ is always
asymmetric: the Kondo peak is slightly shifted toward the positive (negative)
frequency part for $\delta>0$ ($\delta<0$), which leads to a positive
(negative) peak in $L$ at $T \sim T_K$ \cite{Costi:2010, Molenkamp:2005}: As
shown in \figref{fig:MF:em0}(g) $L$ has a positive peak at $T \sim T_K \sim
10^{-4}$ since $\delta > 0$.
This asymmetry can be understood in terms of the slope of the spectral density
at the Fermi level $\epsilon_F$. From the Luttinger theorem \cite{Hewson:1993}, the derivative of
the spectral density is given by
\begin{equation}
  \left.
    \frac{\partial{\varA_{\mu}(\epsilon)}}{\partial\epsilon}
  \right|_{\epsilon=\epsilon_{F}}
  =
  \frac{\sin{(2\avg{n_{\mu}}\pi)}\sin^{2}{(\avg{n_{\mu}}\pi)}}%
  {\pi z_{\mu}\Gamma^{2}},
\end{equation}
where $z_{\mu} = 1/(1-\partial_{\epsilon}\Sigma^{R}_{\mu}(\epsilon_{F}))$,
$\Sigma^{R}_{\mu}$ is the real part of the retarded self-energy per spin and
$\avg{n_{\mu}}$ is the dot occupation per spin.  For $\delta>0$ ($\delta<0$),
$\avg{n_{\mu}}$ is less (larger) than 1/2 so the derivative becomes positive
(negative) and, accordingly, the Kondo resonance is slightly shifted toward the
positive (negative) frequency.
%
%
At larger frequencies, the side peaks in $A_\mu(\omega)$ are responsible for
the two peaks in $L$ at $T \sim |\epsilon_d|$ and $|\epsilon_d+U|$ whose signs
are opposite.  Hence, in the Kondo regime, $L$ experiences two sign changes as
$T$ increases from $T_K$ to high temperatures as predicted in
Ref.~[\onlinecite{Costi:2010}] and reproduced in our calculations.

In the presence of the MF, on the other hand, an additional peak, whose sign is
opposite to that of the Kondo-induced peak, occurs at $T \sim \Gamma_m <
T_K$. It is surely due to the particle-hole asymmetry of the half-dip in
$A_\down(\omega)$ whose width is $\Gamma_m$ [see \figref{fig:MF:em0}(c)]. This
asymmetry also originates from that of the Kondo resonance peak. The
destructive interference responsible for the dip is stronger at the frequency
side in which the Kondo peak is located. As a result, the dip is also slightly
shifted toward the same frequency side as the Kondo peak, leading to the
asymmetry of the dip in the spectral density and the observed peak in $L$.
As $t_{m}$ increases, the half-dip is widened and eventually trims the Kondo
resonance peak when the dip width is comparable to the Kondo peak width [see
\figref{fig:MF:em0}(c)]. Hence, the MF-induced peak in $L$ disappears when
$\Gamma_m \sim T_K$.

\subsection{Strong-coupling regime, $\Gamma_{m} > T_{K}$}

In the strong-coupling regime, the Kondo effect still survives but is modified
as follows: First, the Kondo peak in $A_\up(\omega)$ shifts with increasing
$t_m$ [see \figref{fig:MF:em0}(b)]. The peak moves toward positive (negative)
frequencies for $\delta>0$ ($\delta<0$). It is attributed to the induced Zeeman
splitting $\Delta_Z$ caused by the Coulomb interaction and the broken
time-reversal symmetry in the TSW \cite{Minchul13}. While its unnormalized
value for $t_m \ll |\epsilon_d|, \epsilon+U$ is given by
\begin{equation}
  \label{eq:delta}
  \Delta_{Z}^{(0)}
  =
  -\frac{t^{2}_{m}}{2}
  \frac{\delta}{(\epsilon_{m}+\delta-\epsilon_{d})(\epsilon_{m}-\epsilon_{d})}
\end{equation}
for a general value of $\epsilon_m$, in fact the renormalization due to the
coupling to the leads makes it larger than $\Delta_Z^{(0)}$ in magnitude
\cite{Minchul13}. On the other hand, $A_\down(\omega)$ is not affected by the
induced Zeeman splitting because it is directly coupled to the zero-energy MF
and pinned to the Fermi level. The height of $\pi\Gamma A_\down(\omega)$ is
fixed at $1/2$ [see \figref{fig:MF:em0}(d)].
Since the Kondo peak in $A_\up(\omega)$ is shifted, the linear conductance at
low temperatures is smaller than the weak-coupling-limit value, $3e^2/2h$ and
decreases to $e^{2}/2h$ as $t_m$ increases [see
\figref{fig:MF:em0}(f)].

The effect of the QD-MF coupling in the thermoelectrical conductance is the
shift of the Kondo-induced peak in $L$ with increasing $t_m$ [see
\figref{fig:MF:em0}(h)]. The shift is obviously due to that of the Kondo peak
in $A_\up(\omega)$ by the induced Zeeman splitting explained above. Hence, the
position of the Kondo-induced peak in $L$ should follow the induced Zeeman
splitting: $T \sim \Delta_Z$.

\subsection{SF case}

\begin{figure}
  \includegraphics[width=8.5cm]{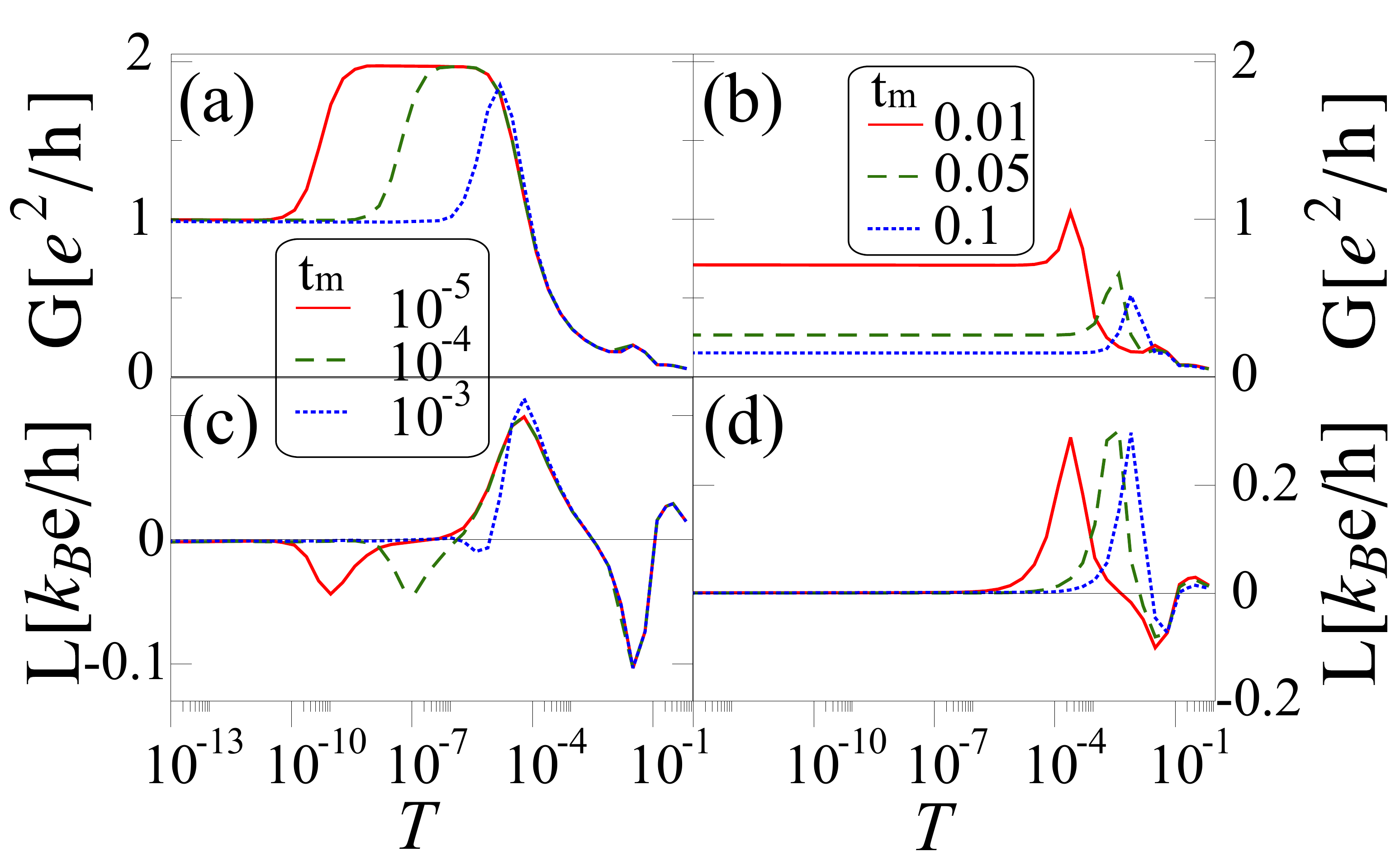}
  \caption{(Color online) Electric/thermoelectric transport features in the SF
    case for $\epsilon_m = 0$. Left and right panels correspond to the weak-
    ($\Gamma_m<T_K$) and strong-coupling ($\Gamma_m>T_K$) regimes,
    respectively: [(a),(b)] electrical conductance and [(c),(d)]
    thermoelectrical conductance for annotated values of $t_m$. Here we have
    used the same values of the parameters as in \figref{fig:MF:em0}.}
  \label{fig:SF:em0}
\end{figure}

Now one should ask whether the observed MF-induced features in $L$ are the
genuine effects of the Majorana physics or not. For that purpose, we have
repeated the same calculations by using the Hamiltonian, \eqnref{eq:H:SF} with
a zero-energy spin-polarized level ($\epsilon_m=0$). First, we have obtained
the qualitatively same spectral densities as in the MF case: $A_\up(\omega)$
features the Kondo resonance peak at $\omega=0$ in the weak-coupling regime and
its peak moves due to the effective Zeeman splitting in the strong-coupling
regime. $A_\down(\omega)$ also exhibits the dip structure due to the anti-Fano
resonance by the side-coupled spin-polarized level. One and only difference
between two cases is that the dip in the SF case is a full dip so that
$A_\down(\omega=0) = 0$.
As a result, the low-temperature value of the linear conductance, which is
$e^2/h$ in the weak-coupling regime, is decreased compared to the MF case [see
\figsref{fig:SF:em0}(a) and \ref{fig:SF:em0}(b)]. However, the other features
of the linear conductance is the same as those in the MF case.
The thermoelectrical conductance also exhibits the same qualitative behavior as
that of the MF case [see \figsref{fig:MF:em0}(c) and \ref{fig:MF:em0}(d)]: the
appearance of the MF-induced peak at $T\sim\Gamma_m$ in the weak-coupling
regime and the shift of the Kondo-induced peak at $T\sim\Delta_Z$ in the
strong-coupling regime.
Hence, for $\epsilon_m = 0$, while the two cases exhibit different behaviors of
the electrical conductance, the measurement of the thermoelectrical conductance
cannot distinguish the MF state from the ordinary spin-polarized state.

\section{Short TSW Case: $\epsilon_m \ne 0$}

The overlap between Majorana states depends on the relative length between the
TSW length and the localization length of the Majorana states. Hence, by
changing the relative length (for example, the TSW length by tuning the
position-dependent gate potential or the MF localization length by controlling
the superconducting gap and/or the gate potential), the overlap amplitude
$\epsilon_m$ can be varied. Similarly, the level of the ordinary spin-polarized
state can be tuned by electric or magnetic methods. In this section, we examine
whether the MF can display a genuine thermoelectric effect for finite
$\epsilon_m$, which cannot be reproduced by the SF level.

\subsection{Weak-coupling regime, $\Gamma_{m} < T_{K}$}

\begin{figure}
  \includegraphics[width=8.5cm]{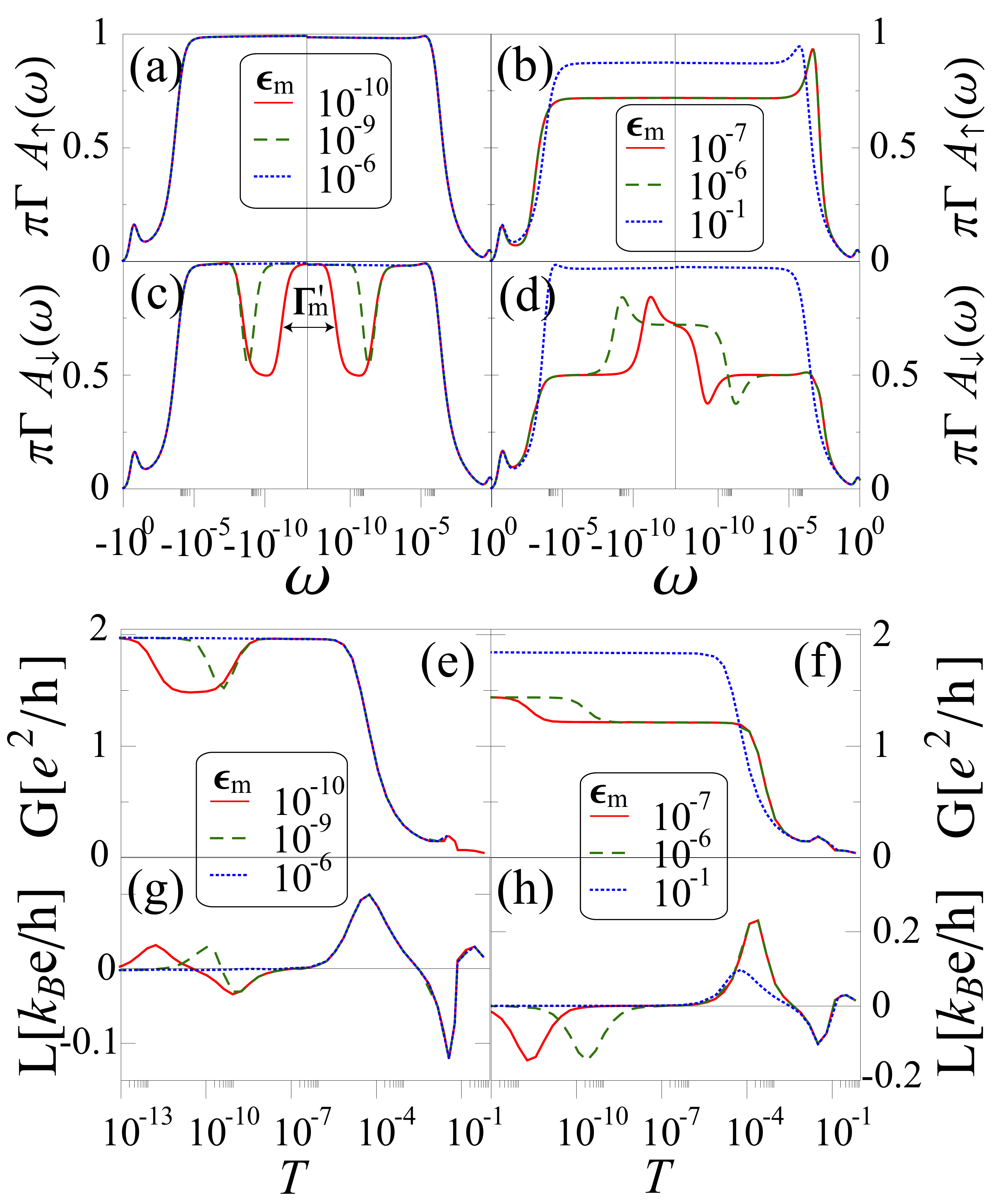}
  \caption{(Color online) Spin-resolved spectral densities and
    electric/thermoelectric conductances in the MF case for $\epsilon_m \ne
    0$. Left and right panels correspond to the weak- ($t_{m} = 10^{-5}$) and
    strong-coupling ($t_{m} = 10^{-2}$) regimes, respectively: [(a),(b)] QD
    spin-$\up$ spectral density, [(c),(d)] QD spin-$\down$ spectral density,
    [(e),(f)] electrical conductance, and [(g),(h)] thermoelectrical
    conductance for annotated values of $\epsilon_m$. Here we have used the
    same values of the parameters as in \figref{fig:MF:em0}.}
  \label{fig:MF}
\end{figure}

The left panels of \figref{fig:MF} show the QD spectral densities, the linear
conductance, and the thermoelectrical conductance in the weak-coupling regime
for $\epsilon_m \ne 0$. As for $\epsilon_m=0$, the Kondo peak in
$A_\up(\omega)$ remains intact irrespectively of the value of $\epsilon_m$ [see
\figref{fig:MF}(a)]. An interesting change occurs in $A_\down(\omega)$: Inside
the half-dip, a central peak is developed for finite values of $\epsilon_m$
[see \figref{fig:MF}(c)], which originates from the interference between the
MFs, $\gamma_1$ and $\gamma_2$. Finite coupling $2\epsilon_{m}$ between
$\gamma_{1}$ and $\gamma_{2}$ broadens the $\gamma_2$ spectral density with the
width $\Gamma^{\prime}_{m}\approx (2\epsilon_{m})^{2}/\Gamma_{m}$ and in turn
the scattering from the resonance level of $\gamma_2$ results in the
destructive interference for $\gamma_1$, forming a full dip in the $\gamma_1$
spectral density. Hence, inside the dip, the anti-Fano resonance which
suppressed the Kondo effect does not happen any longer since $\gamma_1$ which
would interfere with the QD spin has no excitation in that range of energy, and
the Kondo effect is restored. As $\epsilon_m$ increases, the central peak gets
wider and the full Kondo resonance peak is restored when $\Gamma'_m \sim
\Gamma_m$ or $\epsilon_m\sim\Gamma_m$.
The restoration of the Kondo effect at low frequencies is well reflected in the
linear conductance for $T <\Gamma^{\prime}_{m}$ [see \figref{fig:MF}(e)].

The central peak in $A_\down(\omega)$ due to the restoration of the Kondo
effect also has its trace on the thermoelectrical
conductance. \Figref{fig:MF}(g) shows that an additional peak in $L$ at $T \sim
\Gamma'_m$ is formed on top of $L$ in the $\epsilon_m = 0$ case. In fact, the
central peak is also asymmetric like the original Kondo peak. It should be
noted that the dip in the $\gamma_1$ spectral density is symmetric since the MF
always maintains the particle-hole symmetry. So, the asymmetry of the central
peak has the same origin as the Kondo peak, and the signs of the Kondo-induced
peak and the finite-$\epsilon_m$-induced peak in $L$ are always same.
As $\epsilon_m$ increases up to $\sim\Gamma_m$, the half-dip and the inner
central peak in $A_\down(\omega)$ cancels out each other. Therefore, the two
MF-related peaks are merged and disappear completely when $\epsilon_m \sim
\Gamma_m$, making $L\approx0$ for $T \ll T_K$.

\subsection{Strong-coupling regime, $\Gamma_{m} > T_{K}$}

In the strong-coupling regime the finite $\epsilon_m$ removes the
half-fermionic Fano resonance at the Fermi level. As a result, $\pi\Gamma
A_\down(\omega)$, which was pinned to 1/2 at $\omega = 0$ for $\epsilon_m=0$,
becomes larger than 1/2, restoring the Kondo correlation a bit [see
\figref{fig:MF}(d)]. Instead, since the MF states are now energy-split, the
half-value pinning is retained in $\Gamma'_m < |\omega| < \Gamma_m$ for
$\epsilon_m < \Gamma_m$. In the frequency windows $|\omega| < \Gamma'_m$ we
observe in $A_\down(\omega)$ an asymmetric peak (with a small dip in the
opposite side) whose position is approximately at $\omega \approx - {\rm
  sgn}(\delta) \Gamma_m/\pi$. The peak in $A_\down(\omega)$ is located in the
opposite side with respect to the Zeeman-splitting peak in $A_\up(\omega)$ [see
\figref{fig:MF}(b)].  We speculate that the asymmetric peak in
$A_\down(\omega)$ is the outcome of the competition between the induced Zeeman
splitting on spin $\down$ (so the peak would appear at $\omega \approx -
\Delta_Z$) and the half-fermionic pinning at the energy-split MF level. Note
that this central peak has the opposite asymmetry with respect to that in the
weak-coupling regime. Hence, due to this asymmetric peak structure, an
additional peak in $L$ arises at $T \sim \Gamma'_m$, whose sign is opposite to
that of the Kondo-induced peak [see \figref{fig:MF}(h)].
Finally, owing to the increase in $A_\down(\omega)$ at low frequencies, the
low-temperature value of the linear conductance $G$ increases a bit [see
\figref{fig:MF}(f)].

For $\epsilon_m \gg \Gamma_m$, the MF levels which are at $\pm\epsilon_m$ do
not interfere with the Kondo resonant level any longer so that the Kondo
physics in the QD is completely revived \cite{Minchul13}: The peak in $L$ at $T
\sim \Gamma'_m$ disappears.

\subsection{SF case}

\begin{figure}
  \includegraphics[width=8.5cm]{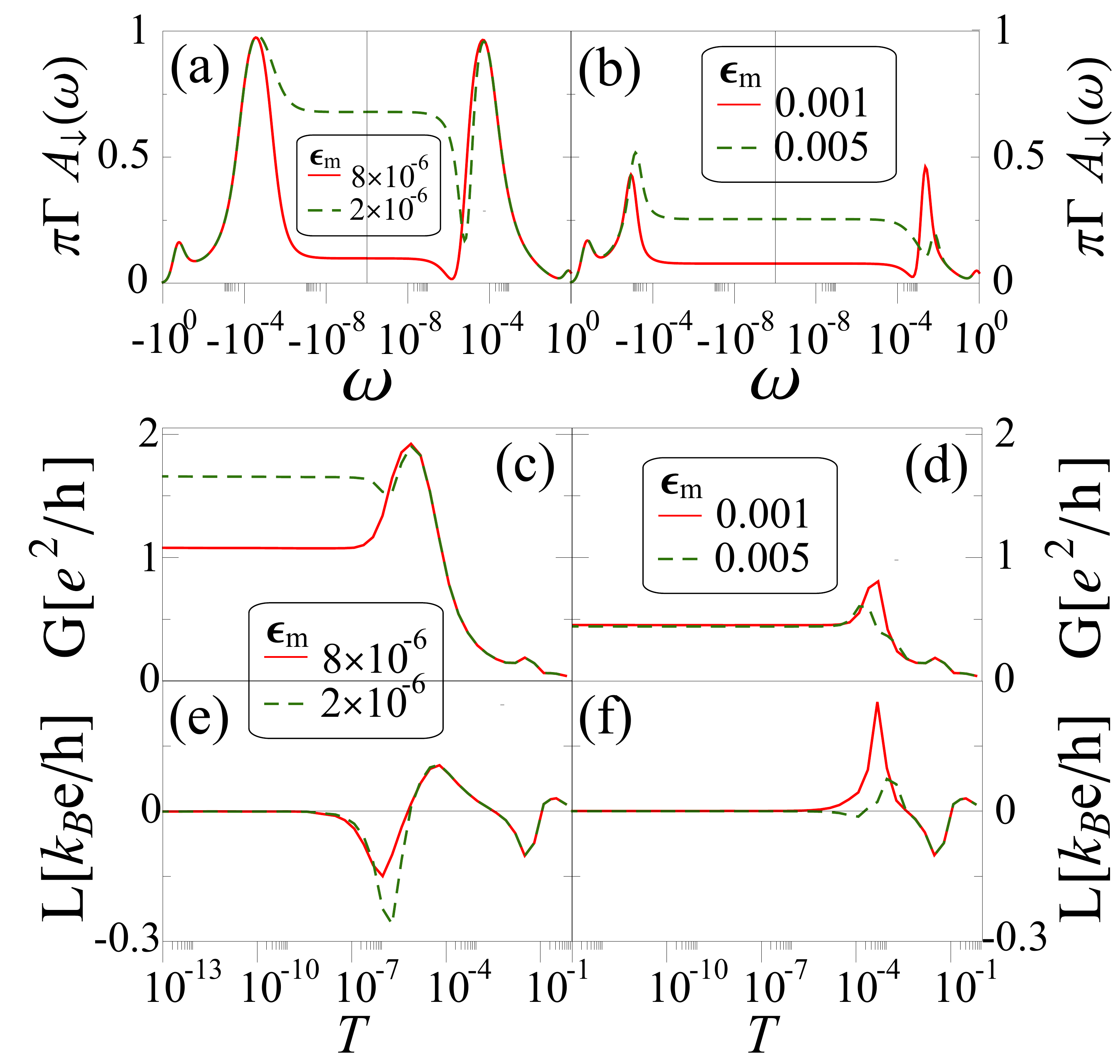}
  \caption{(Color online) Dynamical and electric/thermoelectric transport
    features in the SF case for $\epsilon_m > 0$. Left and right panels
    correspond to the weak- ($t_m\ {=}\ 5\times10^{-4}$) and strong-coupling
    ($t_m = 10^{-2}$) regimes, respectively: [(a),(b)] QD spin-$\down$ spectral
    density, [(c),(d)] electrical conductance, and [(e),(f)] thermoelectrical
    conductance for annotated values of $\epsilon_m$. Here we have used the
    same values of the parameters as in \figref{fig:MF:em0}.}
  \label{fig:SF}
\end{figure}
\begin{figure}
  \includegraphics[width=8.5cm]{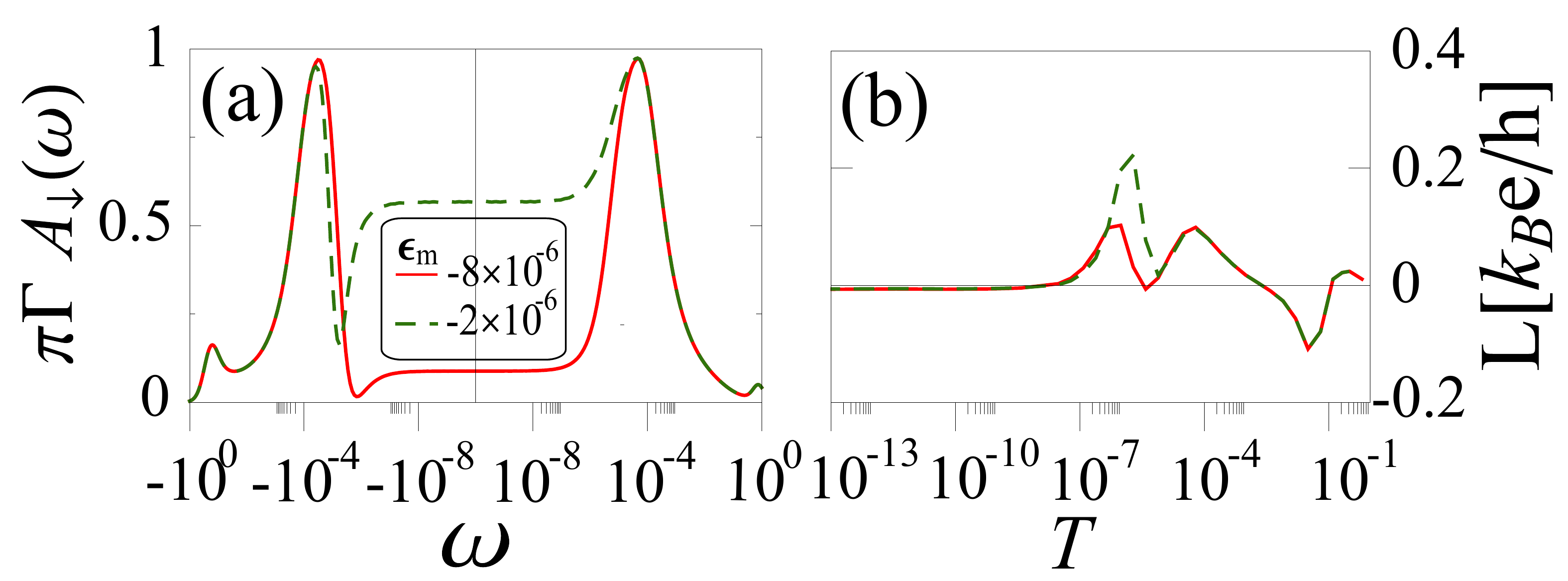}
  \caption{(Color online) (a) QD spin-$\down$ spectral density and (b)
    thermoelectrical conductance in the SF case for $\epsilon_m < 0$ in the
    weak-coupling case. The corresponding values of $\epsilon_{m}$ are
    annotated. Here we use the same values of parameters as in the
    weak-coupling regime of \figref{fig:SF}.}
  \label{fig:SF:neg}
\end{figure}

It is obvious that the additional low-temperature features in $L$ of the MF
case with finite $\epsilon_m$ cannot be reconstructed by the SF case with zero
$\epsilon_m$. For the SF case, the effect of $\epsilon_m$ being finite on the
QD spectral density is quite simple. $A_\up(\omega)$ is not so affected as long
as $|\epsilon_m| < \Gamma_m$, and the full dip in $A_\down(\omega)$ due to the
anti-Fano resonance by the spin-polarized level is formed at $\omega =
\epsilon_m/\hbar$ [see \figsref{fig:SF}(a), \ref{fig:SF}(b) and
\ref{fig:SF:neg}(a)]. The only difference between the weak- and strong-coupling
regimes is whether the dip is so narrow as to be confined inside the Kondo
resonance peak or wide enough to suppress it.
Hence, the electrical and thermoelectrical conductances are qualitatively
similar to those in the $\epsilon_m = 0$ case: compare
\figsref{fig:SF:em0}(a-d) and \ref{fig:SF}(c-f), respectively. Note that for
$\epsilon_m \ne 0$ and in the weak-coupling regime the SF-induced peak in $L$
is formed at $T \sim \epsilon_m$.
For $\epsilon_m < 0$, the dip moves toward the negative frequency [see
\figref{fig:SF:neg}(a)] so that the sign of the SF-induced peak is reversed
[see \figref{fig:SF:neg}(b)].

Therefore, as reflected in \figsref{fig:SF} and \ref{fig:SF:neg}, the SF case
with finite $\epsilon_m$ cannot reproduce the temperature dependence of $L$ in
the MF case with finite $\epsilon_m$. Hence, the observed features of $L$ in
the MF case are genuine effects only the MF can generate. The reason why the MF
and SF cases become distinguishable from each other for $\epsilon_m\ne0$ is
that the MF state is particle-hole symmetric while the SF state is not. In the
MF case the only source of the asymmetry is the Kondo effect while in the SF
case the SF level can also make the system asymmetric. For $\epsilon_m = 0$
this difference did not show up since the levels are at the Fermi level and
particle-hole symmetric.

\section{Gate Dependence of the Thermoelectrical Conductance}

In the previous sections it is shown that the MF adds additional peaks to the
temperature dependence of $L$. Since the MF is particle-hole symmetric, the
finite $L$ is attributed to the asymmetry of the Kondo resonance peak, or in
other words, nonzero $\delta$. Therefore, by tuning the value of $\delta$ or
the gate voltage on the QD, the magnitude and the sign of $L$ can be
controlled. In this section we investigate the gate dependence of $L$ in the
Kondo regime.

Before addressing the weak- and strong-coupling regimes separately, we discuss
the common features seen in both cases. As shown in \figsref{fig:L:gate:wc} and
\ref{fig:L:gate:sc}, the thermoelectrical conductance vanishes at $\delta = 0$,
irrespectively of the values of the other parameters. It is because at $\delta
= 0$ the whole system is particle-hole symmetric even with nonzero
$\epsilon_m$. In addition, $L$ is always odd function upon $\delta \to
-\delta$. It is attributed to the fact that the MF is particle-hole symmetric
and the QD is the only source for the asymmetry. Therefore, as the temperature
(and other parameters) changes, the sign of $L$ changes globally as observed in
the non-interacting case \cite{Lopez:2014}.

\subsection{Weak-coupling regime, $\Gamma_m < T_k$}

\begin{figure}[!t]
  \includegraphics[width=8.5cm]{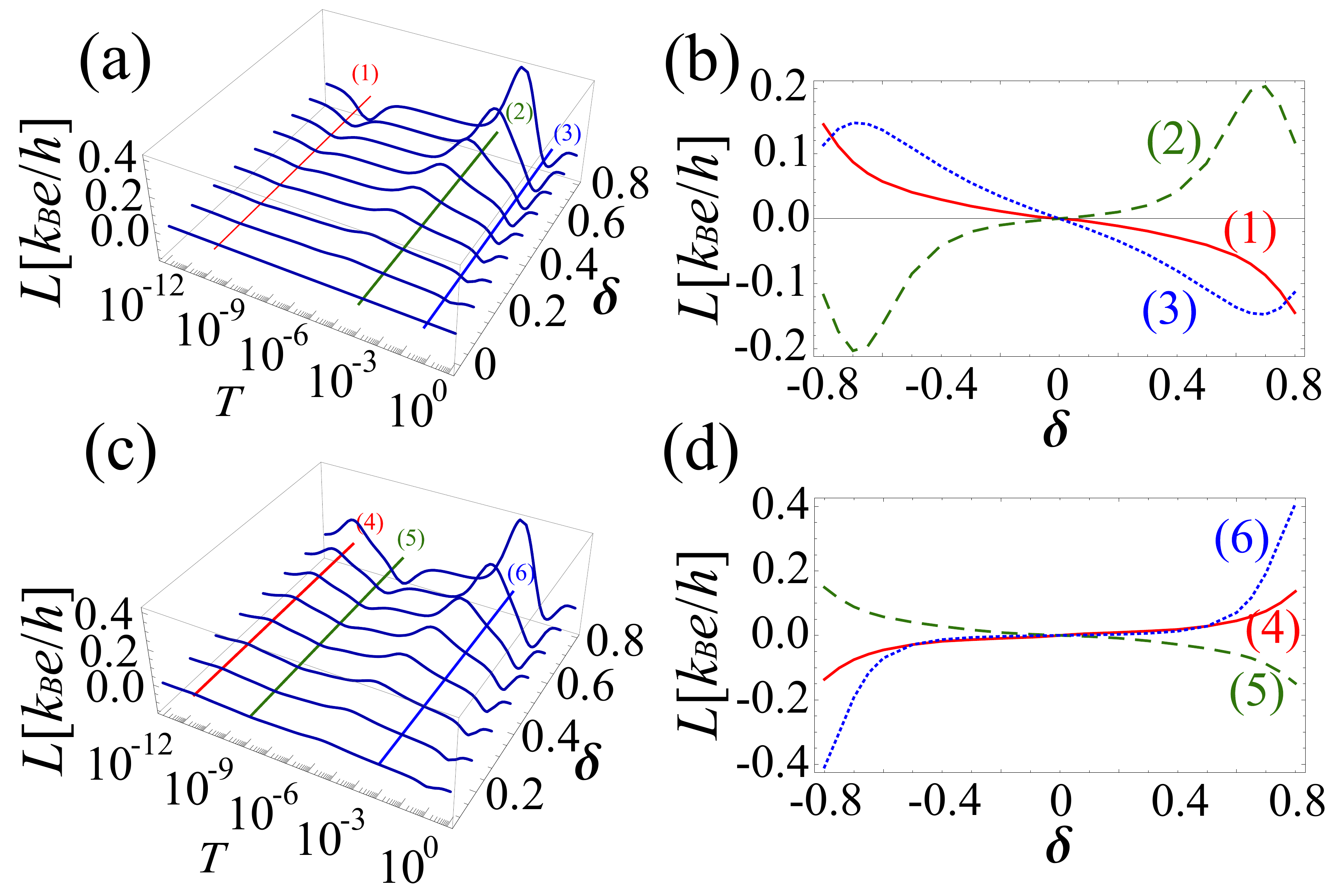}
  \caption{(Color online) Gate dependence of $L$ in the weak-coupling regime
    for [(a),(b)] long TSW case ($\epsilon_{m} = 0$ and $t_{m} = 10^{-6}$) and
    [(c),(d)] short TSW case ($\epsilon_{m} = 10^{-10}$ and $t_{m} = 10^{-5}$).
    In (b) and (d) the thermoelectrical conductances $L$ are drawn as functions
    of $\delta$ at fixed temperatures, or along the straight lines in (a) and
    (c), respectively: (1) MF-induced peak ($T=3.0\times10^{-11}$), (2)
    Kondo-induced peak ($T=1.3\times10^{-4}$), (3) charge-fluctuation-induced
    peak ($T=6.4\times10^{-2}$), (4) restored-Kondo-induced peak
    ($T=3.7\times10^{-12}$), (5) MF-induced peak ($T=1.9\times10^{-9}$), and (6)
    Kondo-induced peak ($T=1.0\times10^{-3}$).}
  \label{fig:L:gate:wc}
\end{figure}

\Figref{fig:L:gate:wc} shows the thermoelectrical conductance as a function of
the temperature $T$ and the asymmetry parameter $\delta$ (or the QD level) in
the weak-coupling regime of both the $\epsilon_m=0$ and $\epsilon_m\ne0$
cases. The main qualitative features of the temperature dependence of $L$
observed in the previous sections do not change with $\delta$, as seen in
\figsref{fig:L:gate:wc}(a) and \ref{fig:L:gate:wc}(c). Instead, the position
and height of the peaks are gradually changed. Owing to the dependence of the
Kondo temperature on the QD level, the Kondo-induced peaks [see curves (2) and
(6) in \figref{fig:L:gate:wc}] move toward the larger temperature as $|\delta|$
increases. However, the MF-related peaks, that is, the MF-induced and
restored-Kondo-induced peaks [see curves (1), (4) and (5) in
\figref{fig:L:gate:wc}] do not move so much upon changing $\delta$ since
$\Gamma_m$ and $\Gamma'_m$ are quite immune to the QD level in the
weak-coupling regime [see \figsref{fig:L:gate:wc}(a) and
\ref{fig:L:gate:wc}(c)]. The heights of the peaks usually increase with
increasing $|\delta|$ [see \figsref{fig:L:gate:wc}(b) and
\ref{fig:L:gate:wc}(d)]. It is because the asymmetry is the larger for the
larger $|\delta|$.

\subsection{Strong-coupling regime, $\Gamma_m > T_k$}

\begin{figure}[!t]
  \includegraphics[width=8.5cm]{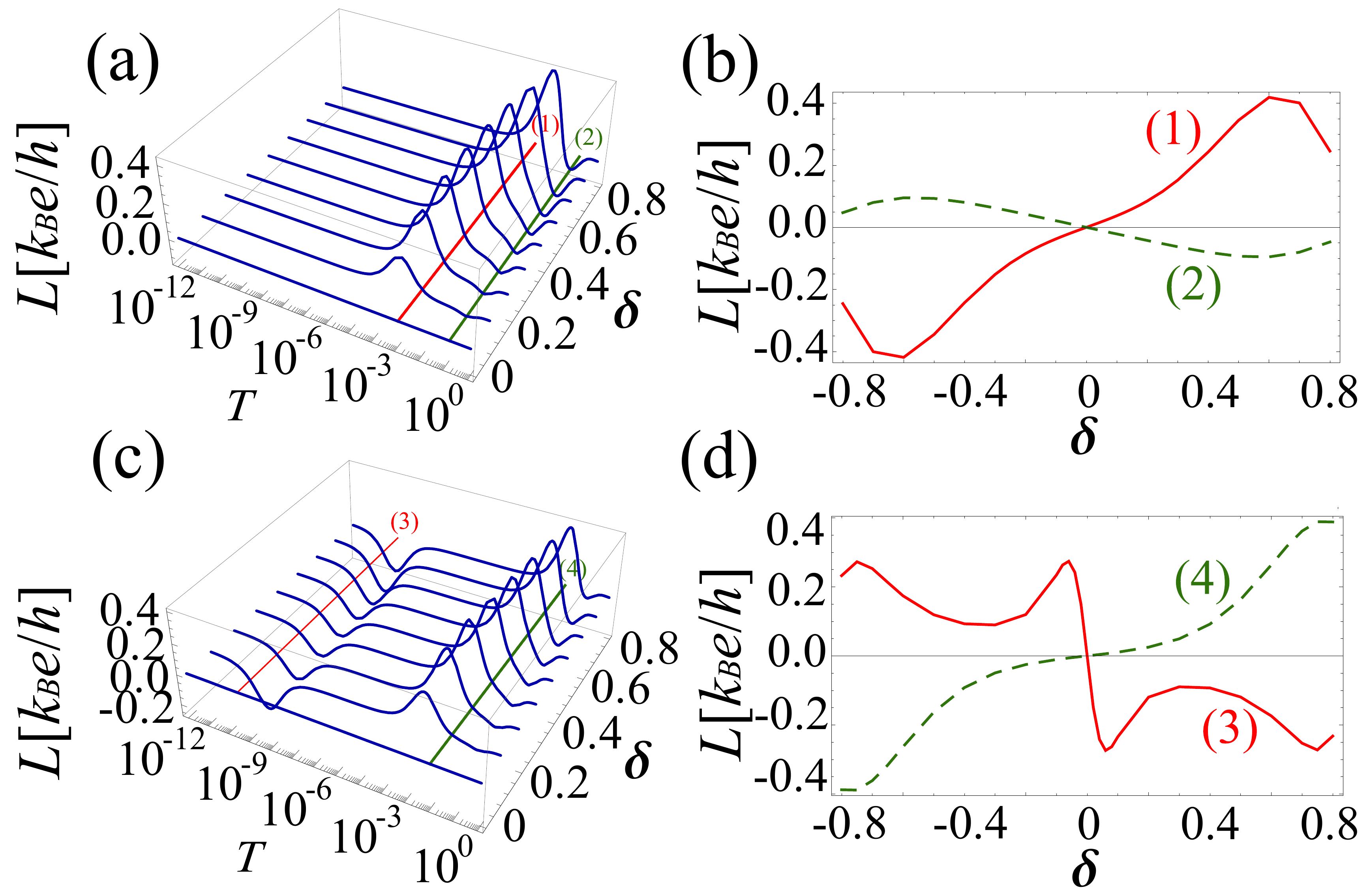}
  \caption{(Color online) Gate dependence of $L$ in the strong-coupling regime
    for [(a),(b)] long TSW case ($\epsilon_{m} = 0$ and $t_{m} = 0.02$) and
    [(c),(d)] short TSW case ($\epsilon_{m} = 10^{-6}$ and $t_{m} = 0.05)$. In
    (b) and (d) the thermoelectrical conductances $L$ are drawn as functions
    of $\delta$ at fixed temperatures, or along the straight lines in (a) and
    (c), respectively: (1) Zeeman-splitting-induced peak ($T=1.0\times10^{-3}$),
    (2) charge-fluctuation-induced peak ($T=1.3\times10^{-1}$), (3)
    restored-Kondo-induced peak ($T=3.0\times10^{-11}$), and (4)
    Zeeman-splitting-induced peak ($T=8.0\times10^{-3}$).}
  \label{fig:L:gate:sc}
\end{figure}

Finally, we consider the gate dependence of the thermoelectrical conductance in
the strong-coupling regime. As in the weak-coupling regime, the main features
of $L$ discussed before are present for all values of $\delta$, as can be seen
in \figref{fig:L:gate:sc}.
For $\epsilon_m=0$ [see \figsref{fig:L:gate:sc}(a) and \ref{fig:L:gate:sc}(b)],
the gate dependence of $L$ in strong coupling regime has similar behavior with
the weak-coupling regime: compare (2) and (3) in \figref{fig:L:gate:wc}(b) with
(1) and (2) in \figref{fig:L:gate:sc}(b), respectively. $L$ increases with
$|\delta|$.
For $\epsilon_m\ne0$ [see \figsref{fig:L:gate:sc}(c) and
\ref{fig:L:gate:sc}(d)], on the other hand, one can observe a non-monotonic
behavior of $L$ with respect to $\delta$ [see (3) in
\figref{fig:L:gate:sc}(d)]. In fact, the peak due to the restored Kondo effect
is shifted as $\delta$ is varied [see \figref{fig:L:gate:sc}(c)]: the peak
first moves toward higher temperature as $|\delta|$ increases from zero and at
some value of $|\delta|$ the shift is reversed, returning back as $|\delta|$
approaches $U$, while the change in the peak position is within the order of
magnitude $\sim \Gamma'_m$. This shift of the peak results in the observed
non-monotonic behavior of $L$ at a fixed temperature. Surely, in the
strong-coupling regime, the Majorana hybridization $\Gamma'_m$ should be
renormalized depending on the value of $\epsilon_d$. While it is not easy to
quantify the renormalization, there is an intuitive way to understand the
non-monotonic behavior: The restore Kondo peak in $A_\down$ in the
strong-coupling regime [see \figref{fig:MF}(d)] is related to the induced
Zeeman splitting $\Delta_Z$ acting on the spin $\down$, as mentioned
before. For small $|\delta|$, $\Delta_Z$ is also small [see \eqnref{eq:delta}]
such that $\Delta_Z < \Gamma'_m$. In this case, the peak follows the induced
Zeeman splitting $\Delta_Z$, explaining the initial increase of the peak
position. However, once $\Delta_Z$ exceeds $\Gamma'_m$, the pinning by the MFs
prevents the peak from moving toward higher temperature further. Instead, as
$|\delta|$ increases, the QD density of states $\rho_{\rm dot}$ increases due
to the resonance levels, and accordingly $\Gamma'_m$ decreases (note that
$\Gamma'_m \propto 1/\Gamma_m \propto 1/\rho_{\rm dot}$). Hence, the peak in
$A_\down$ moves toward lower temperature. This non-monotonic behavior of $L$ is
the outcome of the interplay between the Kondo and Majorana physics so its
experimental observation can be much stronger evidence of the existence of the
MFs.

\section{Conclusion}

We have investigated the thermoelectric properties of strongly correlated
quantum dot side-coupled to the Majorana fermions. The electrical conductance
$G$ and the thermoelectrical conductance $L$ through QD are calculated using
the numerical renormalization group method in both weak- and strong-coupling
regimes. The side-coupled Majorana fermion causes the anti-Fano resonance in
weak-coupling regime and the induced Zeeman splitting in strong-coupling
regime. These two effects considerably modify the shape of the Kondo resonance
peak in the QD spectral densities and eventually transform the structure of
$L$ and $G$. Especially, with a finite overlap between MFs ($\epsilon_{m}\ne
0$), the thermoelectrical conductance $L$ exhibits conspicuous characteristics
-- additional peaks in $L$ and the non-monotonic dependence of $L$ on the gate
voltage -- which cannot be attained by an ordinary fermion. We expect that our
work can act as a guide for finding the unique signal of Majorana fermion in
thermoelectric devices.

\acknowledgements

This work was supported by MINECO Grant No. FIS2011-23526, in part by the Kavli
Institute for Theoretical Physics through NSF grant PHY11-25915, and by the
National Research Foundation of Korea (NRF) grants funded by the Korea
government (MSIP) (No. 2011-0030046).

\bibliographystyle{apsrev4-1}

\begin{thebibliography}{99}
\bibitem{Hasan2010}
  M. Z. Hasan, and C. L Kane, Rev. of Mod. Phys. \textbf{82}, 3045 (2010).

\bibitem{Nayak08}
  C. Nayak, S. H. Simon, A. Stern, M. Freedman, and S. Das Sarma, Rev. Mod. Phys. \textbf{80}, 1083 (2008).

\bibitem{Ettore37}
  E. Majorana, Nuovo Cimento \textbf{14}, 171 (1937).

\bibitem{Kitaev01}
  A. Y. Kitaev, Physics Uspekhi \textbf{44}, 131 (2001).

\bibitem{Wilczek09}
  F. Wilczek, Nat. Phys. \textbf{5}, 614 (2009).

\bibitem{Alicea12}
  J. Alicea, Rep. Prog. Phys. \textbf{75}, 076501 (2012)

\bibitem{Beenakker13}
  C. W. J. Beenakker, Annual Review of Condensed Matter Physics \textbf{4} 113 (2013).

\bibitem{Read91}
  G. Moore and N. Read, Nucl. Phys. B \textbf{360}, 362 (1991).

\bibitem{Ivanov01}
  D. A. Ivanov, Phys. Rev. Lett. \textbf{86}, 268 (2001).

\bibitem{Sarma06}
  S. Das Sarma, C. Nayak, and S. Tewari, Phys. Rev.B \textbf{73}, 220502 (2006).

\bibitem{Ku08}
  L. Fu, and C. L. Kane, Phys. Rev. Lett. \textbf{100}, 096407 (2012).

\bibitem{Linder10a}
  J. Linder, Y. Tanaka, T. Yokoyama, A. Sudb{\o}, and N. Nagaosa, Phys. Rev. Lett. \textbf{104}, 067001 (2010).

\bibitem{Mourik12}
  V. Mourik, K. Zuo, S. M. Frolov, S. R. Plissard, E. P. A. M. Bakkers, and L. P. Kouwenhoven, Science \textbf{336}, 1003 (2012).

\bibitem{Yuval10}
  Y. Oreg, G. Refael, and F. von Oppen, Phys. Rev. Lett. \textbf{105}, 177002 (2010).

\bibitem{Alicea10}
  J. Alicea, Phys. Rev. B \textbf{81}, 125318 (2010).

\bibitem{Lutchyn10}
  R. M. Lutchyn, J. D. Sau, and S. Das Sarma, Phys. Rev. Lett. \textbf{105}, 077001 (2010).

\bibitem{Linder10b}
  J. Linder and A. Sudb\o, Phys. Rev. B \textbf{82}, 085314 (2010).

\bibitem{Potter11}
  A. C. Potter, and P. A. Lee, Phys. Rev. B \textbf{83}, 184520 (2011).

\bibitem{Liu12}
  J. Liu, A. C. Potter, K. T. Law, and P. A. Lee, Phys. Rev. Lett. \textbf{109}, 267002 (2012).

\bibitem{Pientka12}
  F. Pientka, G. Kells, A. Romito, P. W. Brouwer, and F. von Oppen, Phys. Rev. Lett. \textbf{109}, 227006 (2012).

\bibitem{Elsa12a}
  E. Prada, P. San-Jose, R. Aguado, Phys. Rev. B \textbf{86}, 180503(R) (2012).

\bibitem{Lim12}
  J. S. Lim, R. L\'{o}pez, and L. Serra, New J. Phys. \textbf{14}, 083020 (2012).


\bibitem{Deng12}
  M. T. Deng, C. L. Yu, G. Y. Huang, M. Larsson, P. Cardoff, and H. Q. Xu, Nano Lett. \textbf{12}, 6414 (2012).

\bibitem{Heiblum12}
  A. Das, Y. Ronen, Y. Most, Y. Oreg, M. Heinblum, and H. Shtrikman, Nat. Phys. \textbf{8}, 887 (2012).

\bibitem{Churchill12}
  H. O. H. Churchill, V. Fatemi, K. Grove-Rasmussen, M. T. Deng, P. Caroff, H. Q. XU, and C. M. Marcus, Phys. Rev. B \textbf{87}, 241401 (2012).

\bibitem{Finck13}
  A. D. K. Finck, D. J. van Harlingen, P. K. Mohseni, K. Jung, and X. li, Phys. Rev. Lett. \textbf{110}, 126406 (2013).



\bibitem{Chang13}
  W. Chang, V. E. Manucharyan, T. S. Jespersen, J. Nyg\r{a}rd, and C. M. Marcus, Phys. Rev. Lett. \textbf{110}, 217005 (2013).

\bibitem{Eduardo12}
  E. J. H. Lee, X. Jiang, R. Aguado, G. Katsaros, C. M. Lieber, and S. De Franceschi, Phys. Rev. Lett. \textbf{109}, 186802 (2012).

\bibitem{Rok14}
  R. \v{Z}itko, J. S. Lim, R. Lopez, and R. Aguado, arXiv:1405.6084 (2014).

\bibitem{Kells12}
  G. Kells, D. Meidan, and P. W. Brouwer, Phys. Rev. B \textbf{86}, 100503 (2012).

\bibitem{Eduardo13}
  E. J. H. Lee, X. Jiang, M. Houzet, R. Aguado, C. M. Lieber, and S. De Franceschi, arXiv:1302.2611 (2013).

\bibitem{Pikulin12a}
  D. I. Pikulin, J. P. Dahlhaus, M. Wimmer, H. Schomerus, and C. W. J. Beenakker, New. J. Phys. \textbf{14}, 125011 (2012).

\bibitem{Leijnse:2014}
  M. Leijnse, New J. Phys. {\bf 16}, 015029 (2014).

\bibitem{Lopez:2014}
  R. L\'opez, M. Lee, L. Serra, and J. S. Lim, Phys. Rev. B {\bf 89}, 205418 (2014).

\bibitem{Dhar08}
  A. Dhar, Adv. Phys. \textbf{57}, 457 (2008).

\bibitem{Dubi11}
  Y. Dubi and M. Di Ventra, Rev. Mod. Phys. \textbf{83}, 131 (2011).

\bibitem{Molenkamp92}
  L. W. Molenkamp, Th. Gravier, H. van Houten, O. J. A. Buijk, M. A. A. Mabesoone, and C. T. Foxon, Phys. Rev. Lett. \textbf{68}, 3765 (1992).

\bibitem{Dzurak97}
  A. S. Dzurak, C. G. Smith, C. H. W. Barnes, M. Pepper, L. Mart\'{i}n-Moreno, C. T. Liang, D. A. Ritchie, and G. A. C. Jones, Phys. Rev. B \textbf{55}, 10197(R) (1997).

\bibitem{Godjin99}
  S. F. Godijn, S. M\"{o}ller, H. Buhmann, L. W. Molenkamp, and S. A. van Langen, Phys. Rev. Lett. \textbf{82}, 2927 (1999).

\bibitem{Butcher90}
  P. N. Butcher, J. Phys. Condens. Matter \textbf{2}, 4869 (1990).

\bibitem{Coleman05}
  P. Coleman, J. B. Marston, and A. J. Schofield, Phys. Rev. B \textbf{72}, 245111 (2005).

\bibitem{Jacquod10}
  Ph. Jacquod, and R. Whitney, Europhys. Lett. \textbf{91}, 67009 (2010).

\bibitem{Balachandran12}
  V. Balachandran, R. Bosisio, and G. Benenti, Phys. Rev. B \textbf{86}, 035433 (2012).

\bibitem{pekola08}
  B. Kubala, J. K\"{o}nig, and J. Pekola, Phys. Rev. Lett. \textbf{100}, 066801 (2008).

%
%
%
%
%
%
%
%
%
%
%
%
%
%
%
%


\bibitem{Minchul13}
  M. Lee, J. S. Lim, and R. L\'{o}pez, Phys. Rev. B \textbf{87}, 241402 (2013).




%
%
%
%

\bibitem{Leijnse:2011}
  M. Leijnse and K. Flensberg, Phys. Rev. Lett. {\bf 107}, 210502 (2011).

\bibitem{Fu:2008}
  L. Fu and C. L. Kane, Phys. Rev. Lett. {\bf 100}, 096407 (2008).

\bibitem{Sau:2010}
  J. D. Sau, R. M. Lutchyn, S. Tewari, and S. Das Sarma, Phys. Rev. B {\bf 104}, 040502 (2010).

\bibitem{Kitaev:2001}
  A. Y. Kitaev, Physics-Uspekhi \textbf{44}, 131 (2001).
%
\bibitem{Flensberg:2010}
  K. Flensberg, Phys. Rev. B {\bf 82}, 180516 (2010).

\bibitem{Sticlet:2012}
  D. Sticlet, C. Bena, and P. Simon, Phys. Rev. Letts. {\bf 108}, 096802 (2012).

\bibitem{wilson:1975}
  K. G. Wilson, Rev. Mod. Phys. 47, 773 (1975).

\bibitem{Krishna:1980}
  H. R. Krishna-murthy, J. W. Wilkins, and K. G. Wilson, Phys. Rev. B {\bf 21}, 1003 (1980); {\bf 21}, 1044 (1980)

\bibitem{Bulla:2008}
  R. Bulla, T. A. Costi, and T. Pruschke, Rev. Mod. Phys. {\bf 80}, 395 (2008).

\bibitem{Hofstetter:2000}
  W. Hofstetter, Phys. Rev. Lett. 85, 1508 (2000)

\bibitem{campo:2005}
  V. L. Campo and L. N. Oliveira, Phys. Rev. B 72, 104432 (2005).

\bibitem{zitko:2009}
  R. \v{Z}itko and T. Pruschke, Phys. Rev. B 79, 085106 (2009).

\bibitem{Bulla:2001}
  R. Bulla, T. A. Costi, and D. Volhardt, Phys. Rev. B 64 045103 (2001).

\bibitem{Liu:2011}
  D. E. Liu and H. U. Baranger, Phys. Rev. B {\bf 84}, 201308 (2011).

\bibitem{Costi:2010}
  T. A. Costi and V. Zlati\'{c}, Phys. Rev. B {\bf 81}, 235127 (2010).

\bibitem{Molenkamp:2005}
  R. Schelbner, H. Buhmann, D. Reuter, M. N. Klselev, and L. W. Molenkamp, Phys. Rev. Lett {\bf 95}, 176602 (2005).

\bibitem{Hewson:1993}
  A. C. Hewson, {\it The Kondo Problem to Heavy-Fermions} (Cambridge University Press, Cambridge, 1993).


\end{thebibliography}

\end{document}